\documentclass[aps,nofootinbib,twocolumn,prl,floatfix,showpacs]{revtex4} %eqsecnum,
\usepackage{graphicx}
\input epsf

\def\lsim{\mathrel{\raise.3ex\hbox{$<$\kern-.75em\lower1ex\hbox{$\sim$}}}}
\def\gsim{\mathrel{\raise.3ex\hbox{$>$\kern-.75em\lower1ex\hbox{$\sim$}}}}

\def\cmm2{{\,\rm cm^{-2}}}
\def\cm2{{\,{\rm cm}^2}}
\def\cmm3{{\,{\rm cm}^{-3}}}
\def\gcmm3{{\,{\rm g\,cm^{-3}}}}

\def\fun#1#2{\lower3.6pt\vbox{\baselineskip0pt\lineskip.9pt
  \ialign{$\mathsurround=0pt#1\hfil##\hfil$\crcr#2\crcr\sim\crcr}}}

\def\be{\begin{equation}}
\def\ee{\end{equation}}
\def\bea{\begin{eqnarray}}
\def\eea{\end{eqnarray}}

\def\sigv{\langle\sigma v\rangle}

\begin{document}

%\vspace{.2in}
%\title{Inverse Compton Constraints On Dark Matter Annihilation To Leptons}
\title{The Contribution Of Inverse Compton Scattering To The Diffuse Extragalactic Gamma Ray Background From Annihilating Dark Matter}

%\vspace{.2in}
\author{Alexander V.~Belikov$^{1}$ and Dan Hooper$^{2,3}$}
%\vspace{.2in}
\affiliation{$^1$Department of Physics, The
University of Chicago, Chicago, IL~~60637-1433}
\affiliation{$^2$Department of Astronomy \& Astrophysics, The
University of Chicago, Chicago, IL~~60637-1433}
\affiliation{$^3$Center for Particle Astrophysics, Fermi National
Accelerator Laboratory, Batavia, IL~~60510-0500}

\date{\today}
%\smallskip
\begin{abstract}
In addition to gamma rays, dark matter annihilation products can include energetic electrons which inverse Compton scatter with the cosmic microwave background to produce a diffuse extragalactic background of gamma rays and x rays. In models in which the dark matter particles annihilate primarily to electrons or muons, the measurements of EGRET and COMPTEL can provide significant constraints on the annihilation cross section. The Fermi gamma ray space telescope will likely provide an even more stringent test of such scenarios. 
\end{abstract}
\pacs{95.35.+d; %94.20.dv; 
95.85.Pw; FERMILAB-PUB-09-302-A}
\maketitle

\section{Introduction}

Annihilations of dark matter particles are predicted to produce a wide variety of potentially observable products~\cite{review}. These include neutrinos, antineutrinos, electrons, positrons, protons, antiprotons, and photons. Searches for the photons generated in dark matter annihilations have a major advantage over most other indirect detection techniques in that these particles travel essentially unimpeded.  In particular, unlike charged particles, photons are not deflected by magnetic fields, and thus can potentially provide valuable angular information. Furthermore, below approximately $\sim$10 GeV, gamma rays are not attenuated, and can propagate without energy losses over cosmological distances.

The transparency of the universe to gamma rays opens the possibility of observing a diffuse, isotropic background associated with dark matter annihilations from throughout the observable universe. Dark matter particles with weak-scale masses which annihilate to quarks or gauge bosons lead to a diffuse spectrum of energetic photons which peaks at $\sim$1-10 GeV and could potentially be studied with the Fermi gamma ray space telescope (FGST)~\cite{baltz}. This possibility has been studied elsewhere in detail~\cite{cosmological,following}. Recent observations of the cosmic ray positron fraction by PAMELA~\cite{PAMELA}, and the cosmic ray electron (plus positron) spectrum by ATIC~\cite{ATIC} and FGST~\cite{FGSTelectron}, have generated a great deal of interest in dark matter particles which annihilate primarily to leptons~\cite{leptons,sommerfeld,us}. The energetic electrons produced in such annihilations rapidly transfer their energy via inverse Compton scattering into lower energy photons, such as those constituting the cosmic microwave background. The final product of this process is a diffuse extragalactic background of lower energy gamma rays and x rays.

In this paper, we study the extragalactic diffuse gamma ray background resulting from the inverse Compton scattering of energetic electrons produced in dark matter annihilations. We find that if the annihilation cross section (or equivalently, the boost factor) is chosen to generate the signal observed by PAMELA, ATIC, or FGST, a very bright isotropic gamma ray background is predicted. Existing measurements from COMPTEL~\cite{comptel} and EGRET~\cite{egret} can be used to constrain the dark matter annihilation cross section. Observations by FGST will provide a powerful test of dark matter annihilating to electrons or muons.

\section{The Extragalactic Diffuse Gamma Ray Background From Dark Matter Annihilations}

The dark matter annihilation rate per volume at a redshift $z$ is given by
\begin{equation}
R(z) = \int^{\infty}_{M_{\rm min}} \frac{dn}{dM}(M,z)(1+z)^3 dM \frac{\sigv}{2 m^2_X} \int \rho^2(r,M) \,4 \pi r^2 dr, 
\end{equation}
where $dn/dM$ is the differential comoving number density of dark matter halos of mass $M$, $\sigv$ is the dark matter annihilation cross section, $m_X$ is the mass of the dark matter particle, and $\rho$ is the density of dark matter in a halo as a function of the distance from the center of the halo, $r$. 

The number density of dark matter halos of a given mass as a function of redshift is given by 
\begin{equation}
\frac{dn}{dM}(M,z) = \frac{\rho_M}{M} \frac{\ln\sigma^{-1}(M,z)}{dM} f(\sigma^{-1}(M,z)),
\end{equation}
where $\rho_M$ is the average matter density, $\sigma(M,z)$ is the variance of the linear density field, and $f(\sigma^{-1})$ is the multiplicity function. The redshift and cosmology dependence is contained in $\sigma(M,z)$, which can be defined in terms of the matter power spectrum, $P(k)$, and top-hap function, $W(k, M) = (3/k^3 R^3)[\sin(kR) - kR \cos(kR)]$, where $R = (3M/4\pi\rho_m)^{1/3}$,
\begin{equation}
\sigma^2 (M,z) = D^2(z) \int^\infty_0  P(k) W^2(k, M) k^2dk.
\end{equation}
In determining the cold dark matter power spectrum~\cite{Bardeen}, we adopt $\Omega_bh^2 = 0.02267$, $\Omega_c h^2 = 0.1131$, $\Omega_\Lambda = 0.726$, and $h=0.705$, as measured by WMAP~\cite{wmap}. 
%The power spectrum $P(k) \propto k^{n_s} T^2(k)$, where the spectral index $n_s$ comes to play.
The growth function, $D(z)$, is the linear theory growth factor, normalized to unity at $z=0$~\cite{ECF}.
We use the ellipsoidal (Sheth-Tormen) form of the multiplicity function~\cite{ST}, 
\begin{equation}
f(\sigma) = A\frac{\delta_{sc}}{\sigma}\left(1 + \left(\frac{\sigma^2}{a\delta^2_{sc}}\right)^p\right)\exp\left(\frac{a\delta^2_{sc}}{\sigma^2}\right),
\end{equation}
where $p = 0.3$, $\delta_{sc} = 1.686$ and $a = 0.75$~\cite{ST2}. We fix $A = 0.3222$ by the requirement that all of the mass resides in halos.
The halo mass function is most sensitive to variations in the cosmological parameters $\sigma_8$ and $n_s$, for which we adopt $n_s = 0.96$ and $\sigma_8 = 0.812$.

To describe the distribution of dark matter particles within halos, we use the Navarro-Frenk-White (NFW) profile~\cite{nfw}, and halo concentrations given by the analytic model of Bullock {\it et al.}~\cite{bullock}. Rather than extrapolating to lower masses, however, we fix the halo concentrations to a constant value below $M=10^5 \, M_{\odot}$ (as in Ref.~\cite{following}). Different choices for the halo profile or concentration could plausibly increase or decrease the annihilation rate and gamma-ray flux, but likely only by a factor of a few or less. 

%%%%

Following Ref.~\cite{following}, we can write the flux of photons from dark matter annihilations throughout the observable universe in the form
\begin{eqnarray}
\frac{d\phi_{\gamma}}{dE_{\gamma, 0}}&=& \frac{\sigv}{8 \pi}\frac{c}{H_0}\frac{\bar{\rho}^2_{X}}{m^2_X}  \\
&\times&\int dz (1+z)^3 \frac{\Delta^2(z)}{h(z)} \frac{dN_{\gamma}}{dE_{\gamma}}\left(E_{\gamma, 0}(1+z)\right) e^{-\tau(z, E_{\gamma, 0})},\nonumber
\end{eqnarray}
where $\bar{\rho}_X$ denotes the average density of dark matter, $\Delta^2(z)$ is the average squared overdensity and $h(z) = \sqrt{\Omega_\Lambda + \Omega_m(1+z)^3}$ with $\Omega_\Lambda$ and $\Omega_m$ being respectively the present fractions of critical density given by dark energy and matter. The function $\tau$ describes the estimated optical depth of the universe to gamma-rays~\cite{tau}. 

The spectrum of gamma-rays per annihilation, $dN_{\gamma}/dE_{\gamma}$, depends on the dominant annihilation channels, and on the annihilating particles' mass. Neutralinos, for example, typically annihilate to final states consisting of heavy quarks or gauge/Higgs bosons. While the decays of such particles produce a spectrum of prompt photons, this spectrum is rather soft. In contrast, the gamma-ray spectrum from a dark matter particle annihilating to leptons can be significantly harder ({\it i.e.} depositing a larger fraction of the energy into photons with energies not far below the dark matter particle’s mass).  For the additional contribution from final state radiation, see Ref.~\cite{kkgamma}.

%part1
In addition to prompt gamma-rays, dark matter annihilations can also produce high energy electrons and positrons.  Such electrons/positrons transfer their energy to the background radiation through inverse Compton scattering. Since background radiation evolves with redshift, the spectrum of inverse Compton photons is calculated as a function of redshift and added to the spectrum of prompt photons. To calculate the prompt photon spectrum from dark matter annihilations, as well as the prompt electron/positron spectrum, we have used the program PYTHIA \cite{pythia}. For details of the inverse Compton spectrum calculation, we refer the reader to the Appendix.
%For an electron or positron of given energy we calculate the spectrum of inverse Compton photons by Monte Carlo method, using the distribution function given in Ref.~\cite{lightman}. To obtain the full spectrum of inverse Compton photons, we convolve the spectral density of inverse Compton photons produced by an electron or positron with a fixed initial energy with electron/positron spectral density at annihilation.
The results in the next section include the spectrum of inverse Compton photons along with the spectrum of prompt photons.

% as well as the prompt electron/positron spectrum,
\section{Results}

\begin{figure*}
\resizebox{7.5cm}{!}{\includegraphics{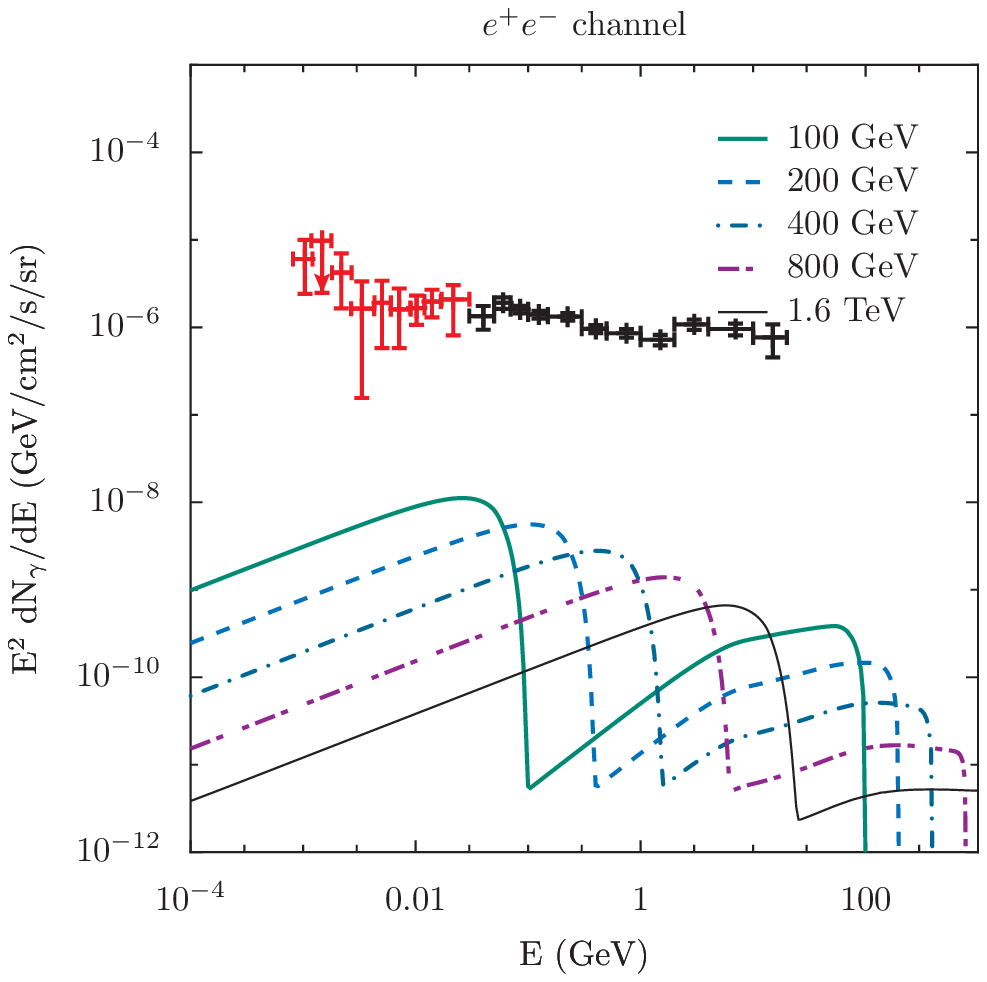}} 
\hspace{0.5cm}
\resizebox{7.5cm}{!}{\includegraphics{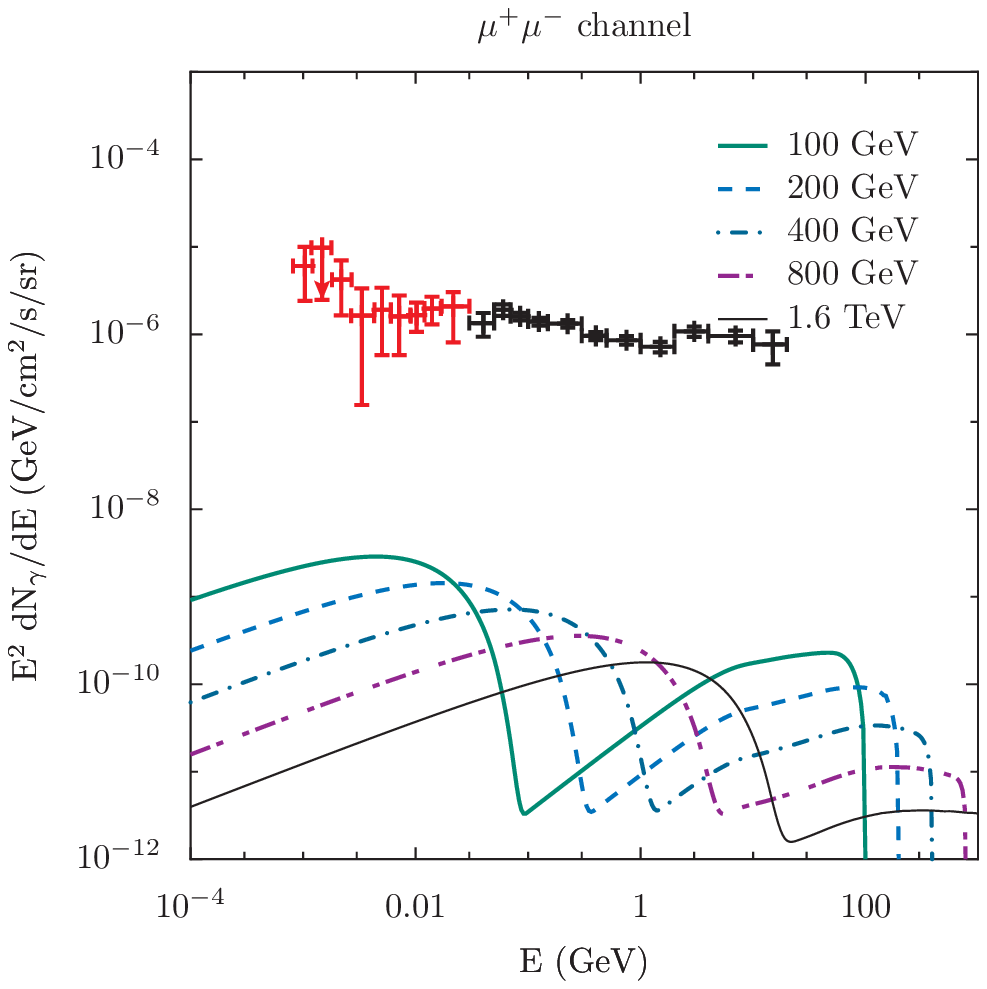}}\\
\vspace{0.4cm}
\resizebox{7.5cm}{!}{\includegraphics{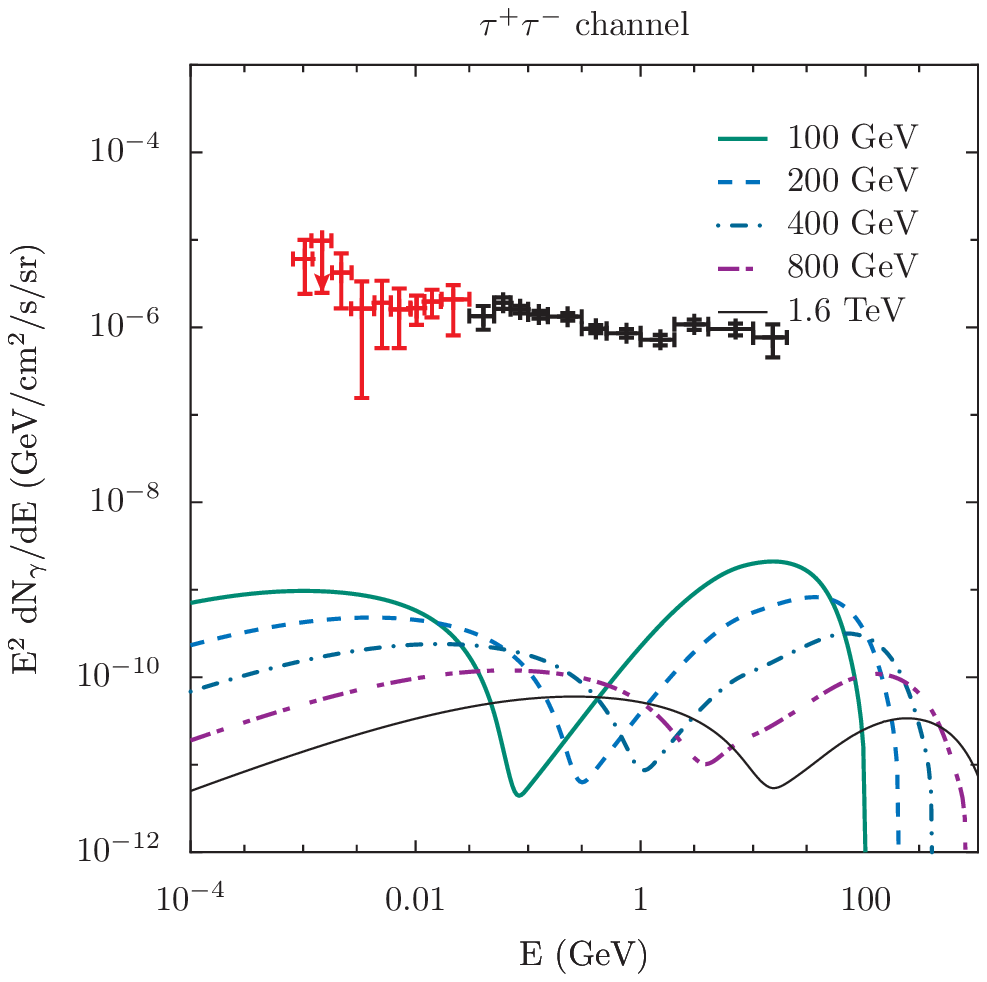}} 
\hspace{0.5cm}
\resizebox{7.5cm}{!}{\includegraphics{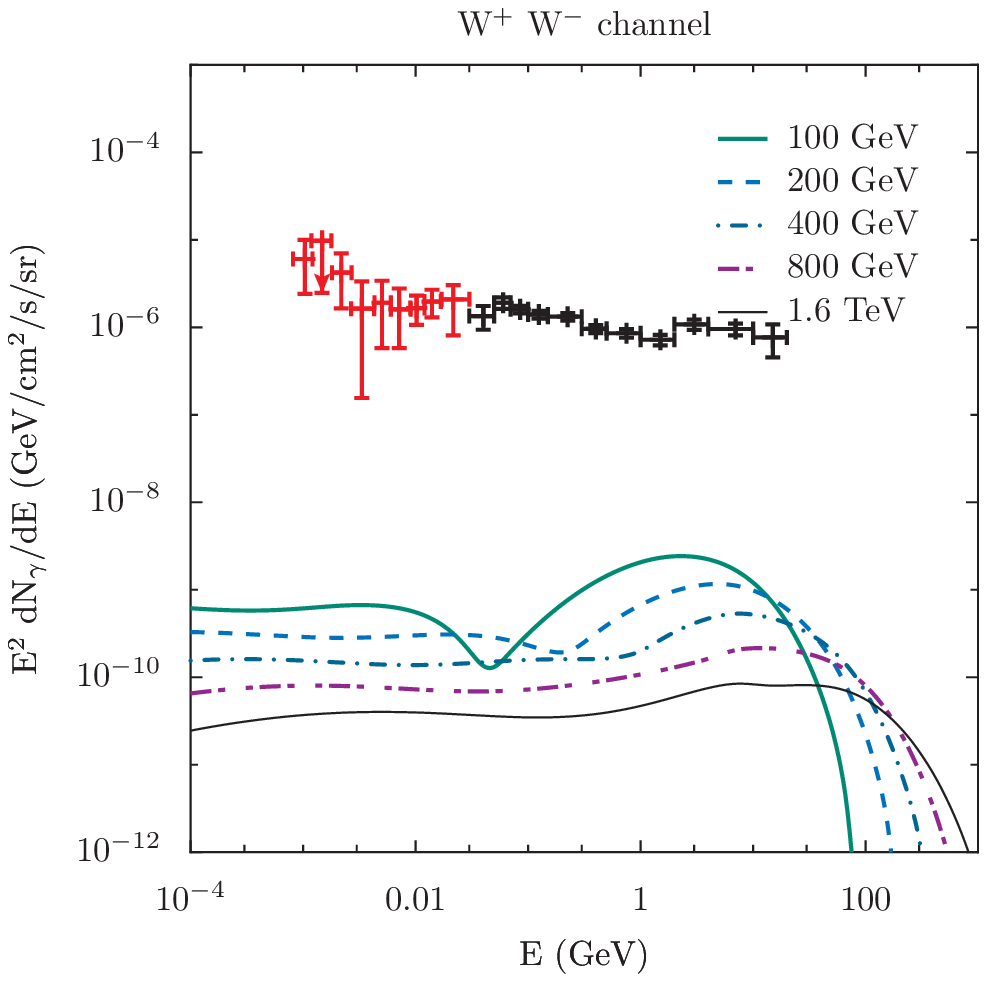}}
\caption{The diffuse extragalactic gamma-ray background from dark matter particles annihilating to $e^+ e^-$, $\mu^+ \mu^-$, $\tau^+ \tau^-$, or $W^+ W^-$. In each frame, the curves denote dark matter masses of 100, 200, 400, 800, and 1600 GeV, from left to right. In each case, we have assumed an annihilation cross section of $\sigv = 3 \times 10^{-26}$ cm$^3$/s (and no boost factor), which is the typical value for a thermal relic. Also shown in each frame are the measurements from the COMPTEL~\cite{comptel} and EGRET~\cite{egret} experiments (on the left and right, respectively).}
\label{fig1}
%\end{center}
\end{figure*}

In Fig.~\ref{fig1}, we show the spectrum of the extragalactic diffuse gamma-ray background from dark matter annihilating to each species of charged lepton ($e^+ e^-$, $\mu^+ \mu^-$, $\tau^+ \tau^-$). For comparison, we also show the results for a dark matter particle annihilating to $W^+ W^-$. In each frame, the curves denote dark matter masses of 100, 200, 400, 800, and 1600 GeV, from left to right. In each case, we have assumed an annihilation cross section of $\sigv = 3 \times 10^{-26}$ cm$^3$/s, which is the typical value for a thermal relic. Also shown in each frame are the measurements from the COMPTEL~\cite{comptel} and EGRET~\cite{egret} experiments.

\begin{figure*}
\resizebox{7.5cm}{!}{\includegraphics{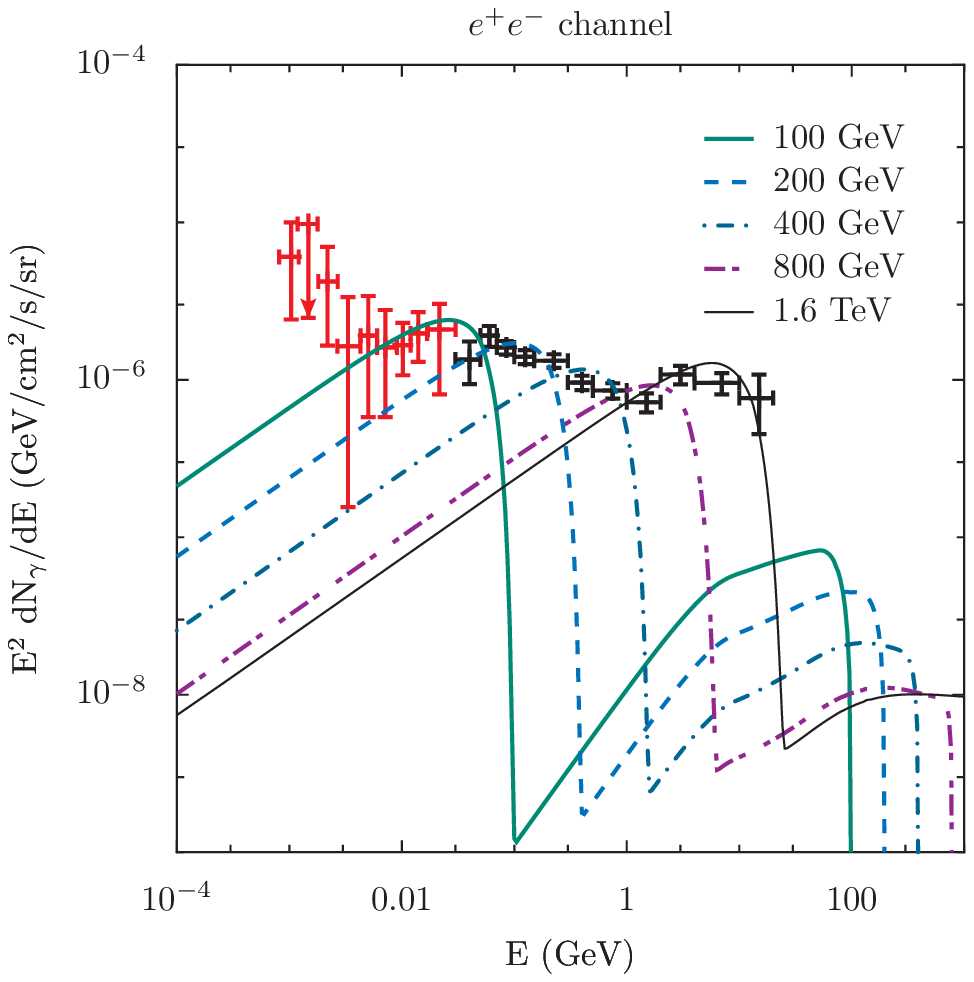}} 
\hspace{0.5cm}
\resizebox{7.5cm}{!}{\includegraphics{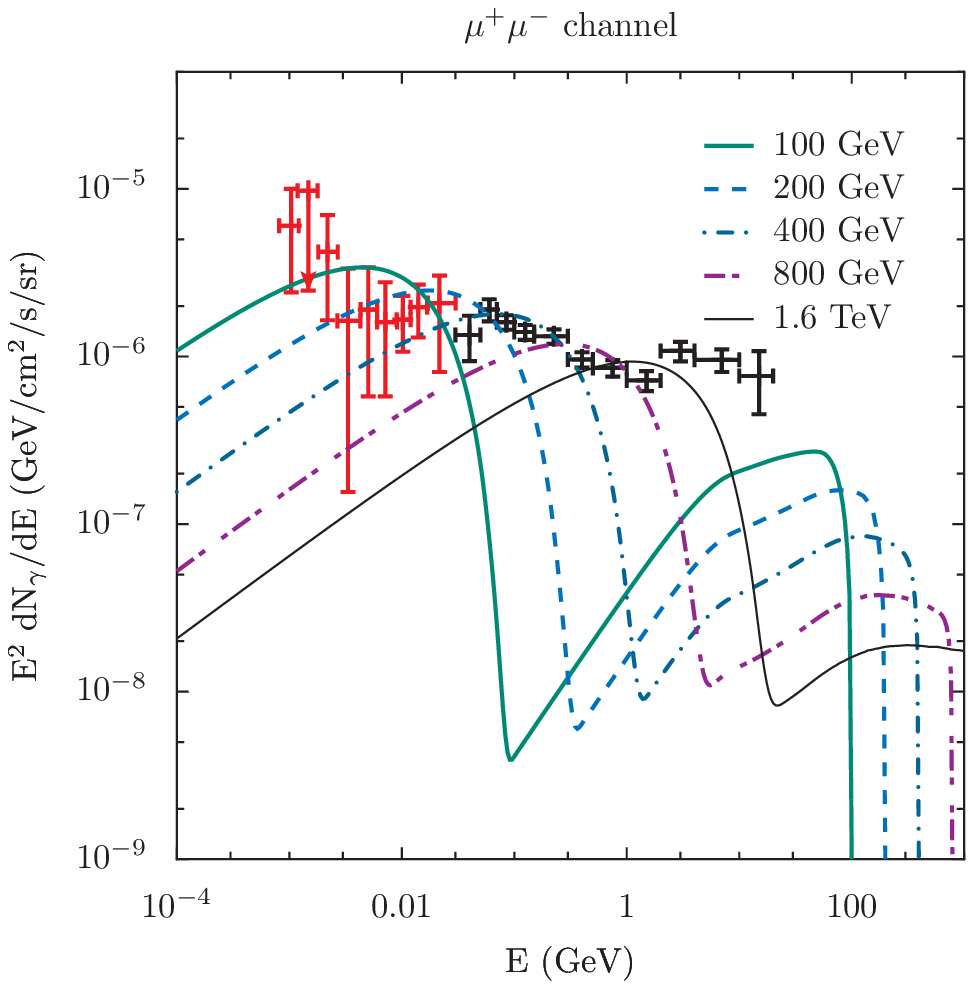}}\\
\vspace{0.4cm}
\resizebox{7.5cm}{!}{\includegraphics{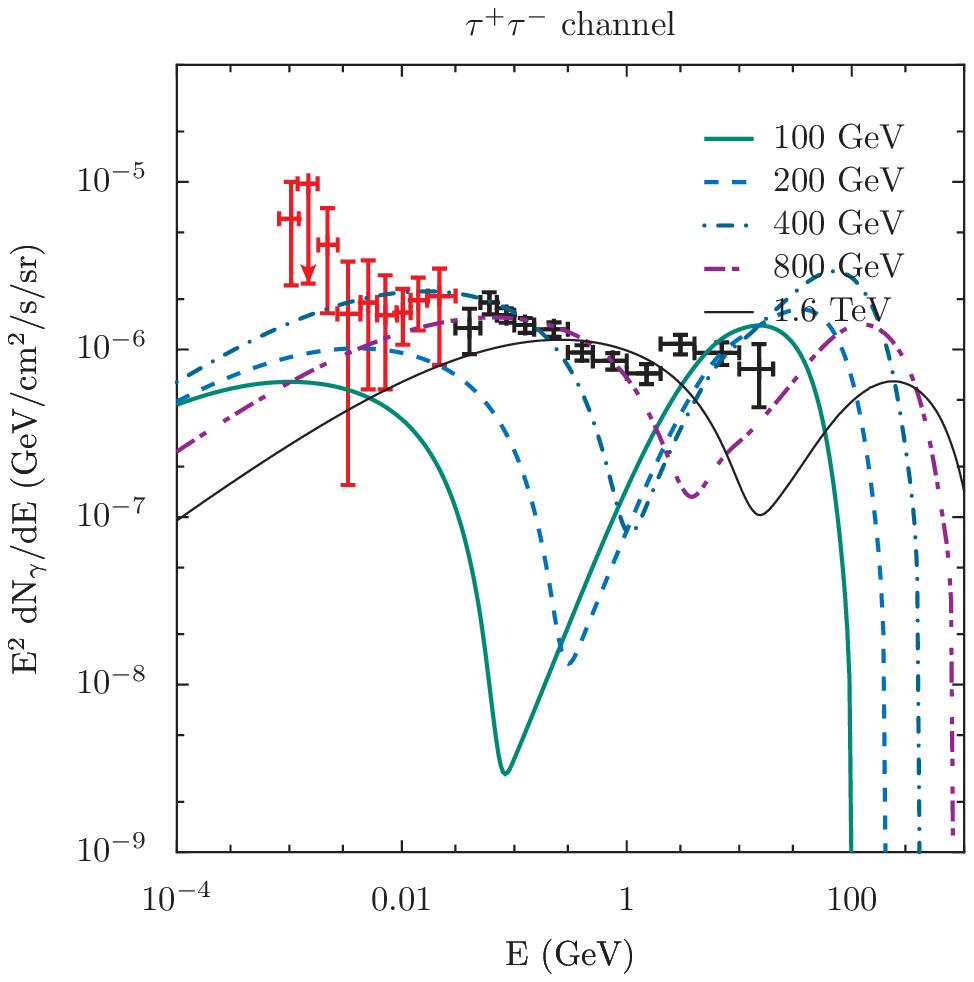}} 
\hspace{0.5cm}
\resizebox{7.5cm}{!}{\includegraphics{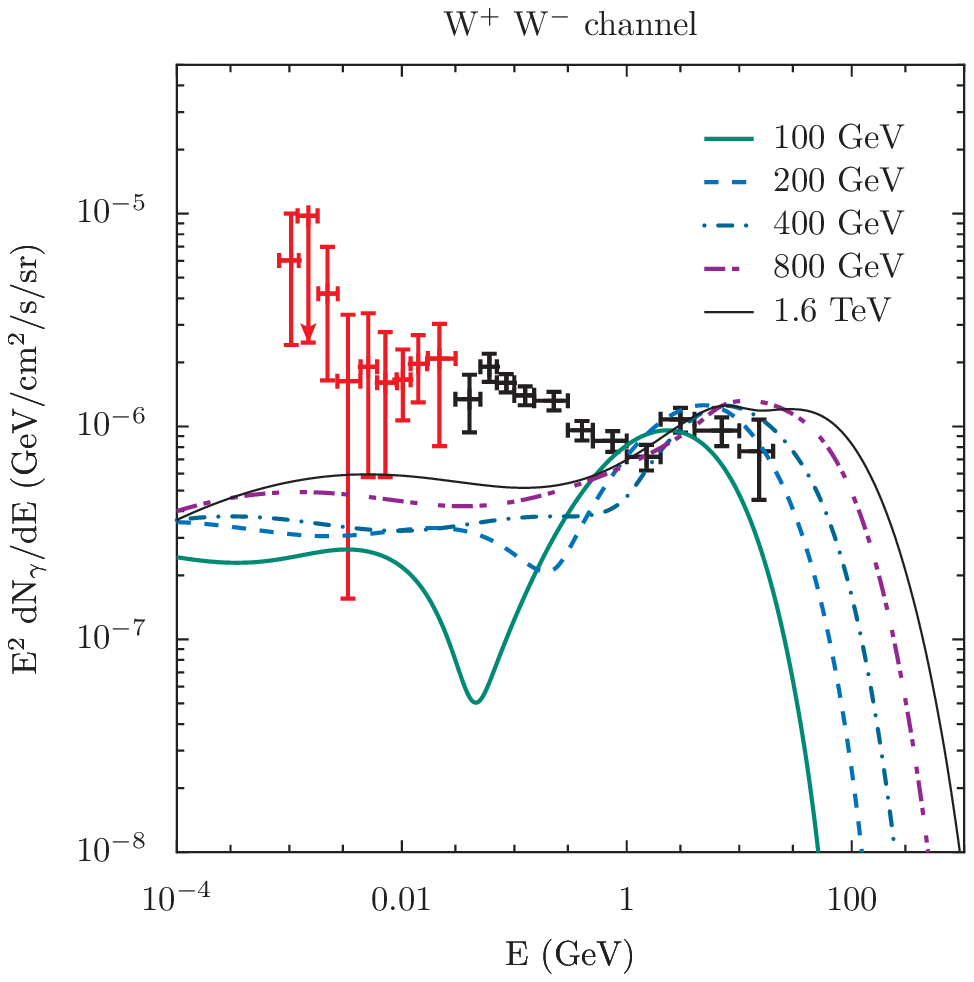}}
\caption{The same as shown in Fig.~\ref{fig1}, but after increasing the annihilation cross section (or boost factor) by the approximate maximum value consistent with the measurements of COMPTEL~\cite{comptel} and EGRET~\cite{egret}. For dark matter annihilating to electrons or taus, the inverse Compton photons provide a far more stringent constraint than prompt emission.}
\label{fig1b}
%\end{center}
\end{figure*}

\begin{figure}
\resizebox{8.5cm}{!}{\includegraphics{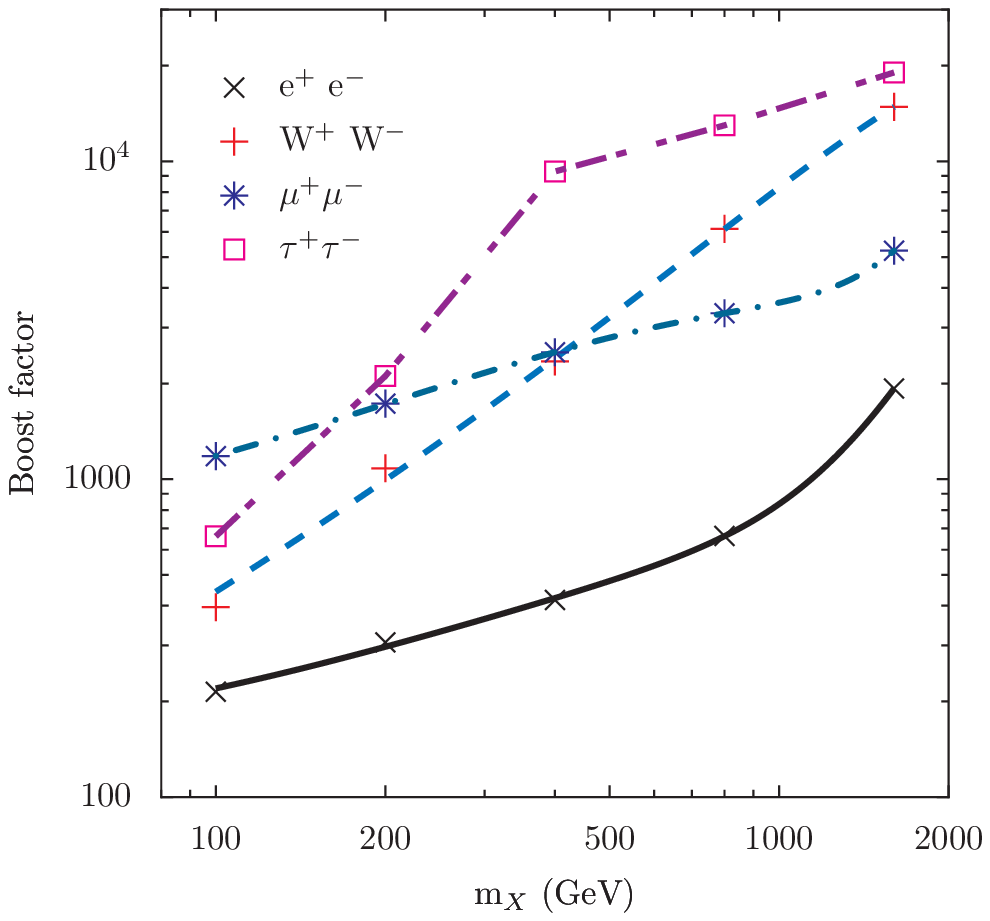}} 
\caption{The maximum value of the boost factor (or equivalently, the maximum enhancement of the annihilation cross section relative to $\sigv = 3 \times 10^{-26}$ cm$^3$/s) consistent with the diffuse isotropic background measured by the COMPTEL~\cite{comptel} and EGRET~\cite{egret} experiments. These constraints are especially stringent for dark matter that annihilates directly to electrons or muons (for WIMP masses above 500 GeV in the latter case).}
\label{fig2}
%\end{center}
\end{figure}

In Fig.~\ref{fig1b}, we again plot the diffuse gamma-ray background from dark matter, but have increased the annihilation cross section (or boost factor) to the maximum value consistent with the measurements of COMPTEL and EGRET. For dark matter particles that annihilate to $W^+ W^-$ (or other varieties of heavy fermions or gauge/Higgs bosons), the strongest constraints typically come from the prompt gamma-ray emission, in contrast to the inverse Compton photons. A 100 GeV (1 TeV) dark matter particle annihilating to $W^+ W^-$ would exceed EGRET's diffuse background measurement if the annihilation cross section were boosted more than $\sim$400 ($\sim7000$) beyond the common thermal estimate ($3 \times 10^{-26}$ cm$^3$/s). Dark matter that annihilates to leptons (especially $e^+ e^-$ or $\mu^+ \mu^-$) deposit a larger fraction of their energy into electrons, and thus ultimately into lower energy inverse Compton photons. In the upper left frame of Fig.~\ref{fig1b}, we find that we cannot increase the annihilation cross section of a 100-500 GeV dark matter particle annihilating to $e^+ e^-$ by more than $\sim$200-500 relative to the thermal estimate without exceeding the measurements of COMPTEL or EGRET. In Fig.~\ref{fig2}, we show these maximum allowed boost factors for each annihilation channel as a function of the dark matter particle's mass. The inverse Compton photons described here may also contribute significantly to the reionization of hydrogen and helium gas at redshifts of $z\sim 6-20$~\cite{reion}.

These results are very interesting within the context of recent observations from PAMELA~\cite{PAMELA}, ATIC~\cite{ATIC}, and FGST~\cite{FGSTelectron}, each of which have been interpreted as possible indications of dark matter annihilations taking place in the Galactic halo~\cite{leptons,us,fgstint}. In order for these signals to be the product of dark matter annihilations, however, the annihilations must proceed largely to charged leptons, and at a considerably higher rate than would be naively predicted for a typical thermal relic distributed smoothly throughout the Milky Way. Thus if one interprets these signals as a product of dark matter annihilation, then a significant diffuse background of gamma rays from the inverse Compton scattering of dark matter annihilation products is also predicted.

As a first example, consider a $\sim 300$ GeV dark matter particle which generates the PAMELA positron excess by annihilating directly to $e^+ e^-$ with a boost factor of $\sim$30-250~\cite{us}. In this case, the inverse Compton photons constitute a distinctive bumplike feature in the diffuse background at $\sim$300 MeV. For a boost factor near the middle of the allowed range ($\sim$1000), this bump will be only a factor of a few below the flux measured by EGRET. If the PAMELA excess is instead the result of the 300 GeV dark matter particle annihilating to muons, the resulting diffuse background will be less than $\sim 10\%$ of that measured by EGRET.

If in addition to the PAMELA positron fraction, one would like to generate the electron spectrum reported by FGST with dark matter annihilations, we must consider a dark matter particle with a $\sim$1-2 TeV mass and which annihilates largely to $\mu^+ \mu^-$. To normalize the annihilation rate to these signals, we must also introduce boost factors ranging from approximately 400-500 for a 1 TeV mass, to 1000 or more for a mass of 2 TeV~\cite{fgstint}. Comparing this to our results, we see that this would lead to a diffuse background constituting about one-fourth of the flux measured by EGRET around $\sim 1$ GeV. Similar conclusions are reached if we consider dark matter that annihilates to muons through a light intermediate state~\cite{fgstint}.

With its superior angular resolution and greater exposure, FGST will study the diffuse extragalactic gamma-ray background between 100 MeV and hundreds of GeV with far greater precision than EGRET. In particular, FGST is expected to resolve and identify many of the individual sources that contribute to the diffuse isotropic background measured by EGRET. It has often suggested that a large fraction of the extragalactic diffuse background measured by EGRET could consist of emission from blazars. As FGST is anticipated to resolve on the order of $10^3$ individual blazars~\cite{prediction} (EGRET, in contrast, only accumulated a catalog of 66 high-confidence blazars over the duration of its mission~\cite{Hartman:1999fc}), the isotropic diffuse background that will be measured by FGST could be significantly smaller than that measured by EGRET~\cite{stecker}. If a significant fraction of the diffuse flux observed by EGRET originates from dark matter annihilations, then FGST may be capable of identifying this signal, especially if a distinctive bumplike inverse Compton feature is present. Alternatively, if FGST resolves a large fraction of the background measured by EGRET, it will be able to place stringent constraints on the dark matter annihilation cross section to electrons or muons.

\section{Summary and Conclusions}

In this article, we have calculated the contribution to the extragalactic, isotropic gamma-ray background from annihilating dark matter particles, including the inverse Compton photons resulting from energetic electrons scattering with the cosmic microwave background. 

Although dark matter particles with typical thermal annihilation cross sections ($\sigv \sim 3 \times 10^{-26}$ cm$^3$/s) are predicted to produce only a relatively small fraction of the isotropic gamma-ray and x-ray backgrounds observed by EGRET and COMPTEL, dark matter particles with larger cross sections (or equivalently, boost factors) could potentially be responsible for much or most of this flux. 

Dark matter scenarios capable of generating the cosmic ray signals observed by PAMELA, ATIC, or the Fermi gamma ray space telescope (FGST) predict dark matter particles which annihilate at a very high rate, and primarily to charged leptons.  In such models, the gamma-ray background from inverse Compton scattering can be very significant, and is constrained by the measurements of COMPTEL and EGRET.

If FGST resolves a significant fraction of the background measured by EGRET into individual sources, it will be capable of placing very stringent limits on dark matter candidates with a high annihilation rate to leptons. If the dark matter annihilates directly to $e^+ e^-$, FGST may also be able to identify the distinctive inverse Compton peak at $E_{\gamma} \sim 300 \, {\rm MeV} \times (m_{X}/300 \, {\rm GeV})^2$ in the diffuse extragalactic gamma-ray spectrum.

\bigskip

While we were in the final stages of completing this paper, Ref.~\cite{profumo} appeared on the LANL preprint archive. Considering that they used the Moore density profile for dark matter halos, which accounts for approximately an order of magnitude flux enhancement over NFW profile, our results are in qualitative agreement. Once the halo profile is taken into account, Ref.~\cite{profumo} finds a flux that is a factor of a few larger than what we have found in our calculation.  The possible sources of this discrepancy may include the multiplicity function of halos, the choice of $\sigma_8$ and $n_s$ and the parametrization of the optical depth~\cite{tau}. We also note that our results for the prompt component of the signal are in very good quantitative agreement with those of Ref.~\cite{following}. This work has been supported by the US Department of Energy, including Grant No. DE-FG02-95ER40896, and by NASA Grant No. NAG5-10842.

\section{Appendix}
%part2
%%%%%%%%%%%%%%%%%%%%%%%%%%%%%%%%%%%%%%%%%%%%%%%%%%%%%%%%%

Relativistic electrons transfer energy to photons in a process known as inverse Compton scattering. 
The rate of scattering of a relativistic electron of energy $E_e = \gamma m_e c^2$ in a gas of monoenergetic photons of energy $E_{\rm rad}$ and density $N_{\rm rad}$ to photons of energy $E_{\rm IC}$ per energy interval $dE_{\rm IC}$ is
\begin{equation}
\frac{dN}{dE_{\rm IC}\,dt} = \frac{3 N_{\rm rad}\sigma_T\,c\,}{4\gamma^2 E_{\rm rad}}(2\epsilon\ln{\epsilon} + \epsilon + 1 - \epsilon^2),
\end{equation}
where $\epsilon = \frac{E_{\rm IC}}{4\gamma^2 E_{\rm rad}}$ and  $\sigma_T$ is the Thompson cross section \cite{lightman}. In particular, electrons can scatter off of photons in the cosmic microwave background (CMB), which in our case might be considered nearly monoenergetic with energy density $U_{\rm CMB} = N_{\rm CMB} E_{\rm CMB} \approx 0.25 \, {\rm eV} /{\rm cm}^3$ at present.
The average energy of the scattered photon is 
\begin{eqnarray}
E_{\rm IC} &=& \frac{4}{3} \, \bigg(\frac{E_e}{m_e}\bigg)^2 \, E_{\rm CMB} \nonumber \\
           &\approx& 3.4 \times 10^{-2} \, {\rm GeV} \,\, \, (1+z) \, \bigg(\frac{E_e}{100 \, {\rm GeV}}\bigg)^2.
\end{eqnarray}
In order to obtain the spectrum of photons produced by a cascade of scatterings of an electron with initial energy $E_e$ we need to average over realizations. Using $\Delta E_e = E_{\rm IC} - E_{\rm CMB} \propto E_e^2$ one can show that $\frac{dN}{dE_{IC}} \propto E_{\rm IC}^{-3/2}$.
Assuming the properties of the photon gas do not change while any given electron is losing most of its energy, the rate of energy transfer is 

\begin{eqnarray}
\frac{dE_e}{dt} &=& -\frac{4}{3} \, \sigma_T \, U_{\rm CMB}(z=0)\, (1+z)^4 \, c \,\bigg(\frac{E_e}{m_e}\bigg)^2 , \nonumber \\
&\approx & -2.3 \times 10^{-13} \, {\rm GeV}/{\rm s} \,\,\, \,(1+z)^4 \bigg(\frac{E_e}{100 \,{\rm GeV}}\bigg)^2
\end{eqnarray}
As a result of this process, a 10 GeV electron at redshift $z=1$ will lose 90\% of its energy to the CMB within about 80 million years. Thus the vast majority of the energy that is deposited via dark matter annihilations into electrons gets almost immediately transferred into the cosmic microwave background (relative to other timescales in the problem). At higher redshifts, or for higher energy electrons, the transfer of energy is even more rapid. 

Generally speaking while an electron/positron is losing its energy the CMB photons are redshifted. The characteristic time of energy loss due to IC scattering for a 100 GeV electron/positron is $\tau_{\rm IC} = \frac{E_e}{dE_e/dt} \approx 3.9 \times 10^{14} \mbox{s} \ll \tau_H \approx 4.35 \times 10^{17} \mbox{s}$, which is much less than the Hubble time, allowing us to neglect the redshifting of CMB photons.  The spectral density of photons produced by an electron/positron of a given energy $\frac{dN}{dE_{\rm IC}}(E_{\rm IC}, E_e)$ is then calculated by the Monte Carlo method.

To obtain the spectral density of inverse Compton photons for a given redshift per annihilation, one convolves the electron/positron density spectral density per annihilation with the spectral density of IC photons:
\begin{equation}
\frac{dN}{dE_{IC}} = \int\limits^{M_\chi}_{m_ec^2} dE_e \frac{dN}{dE_{IC}}(E_{IC}, E_e)\frac{dN}{dE_{e}}(E_e)
\end{equation}

The photon spectral density is then normalized to carry the same energy density as the electrons/positrons per annihilation. 

%%%

\end{document}